\documentclass[12pt]{article}
\usepackage{epsfig}
\usepackage{amsmath}

\input{tcilatex}

\begin{document}

\title{A new $C$-integrable limit of second harmonic generation equations}
\author{A.M.~Kamchatnov $^{a}$ and M.V.~Pavlov $^{b}$ \\
$^{a}$Institute of Spectroscopy, Russian Academy of Sciences,\\
Troitsk, Moscow Region 142190, Russia\\
$^{b}$ Moscow Technical University of Aviation and Technology,\\
Orshanskaya, 3, Moscow, Russia\\
}
\maketitle

\begin{abstract}
A new $C$-integrable limit of the second harmonic generation equations is
found. The corresponding general solution is given in an explicit form.
Connection of this problem with the modified Liouville equation is discussed.
\end{abstract}


Second harmonic generation (SHG) is a well-known method of experimental
investigations in nonlinear optics. Under idealized condition of only one
space dimension this process is described by the equations (see, e.g., \cite%
{AVC}) 
\begin{equation}  \label{eq1}
q_{1\chi}=-2q_2q_1^*,\qquad q_{2\tau}=q_1^2,
\end{equation}
where differentiation with respect to characteristic variables $\chi$ and $%
\tau$ is indicated by subscripts, the asterisk denotes complex conjugation,
and $q_1$ and $q_2$ are slowly varying complex electric field amplitudes of
two electromagnetic waves with carrier frequencies $\omega_1$ and $%
\omega_2=2\omega_1$, respectively. The SHG equations (\ref{eq1}) are shown
to be ``$S$-integrable'' \cite{Kaup} what permitted one to find some
particular solutions describing transformation of waves. However,
application of the inverse scattering transform method to the Cauchy problem
meets some peculiar difficulties and is not fully developed yet. Therefore,
other approaches to solving the SHG equations are of considerable interest.

As it was noticed long ago \cite{BS}, in the case of pure amplitude
modulation, when $q_1$ and $q_2$ are real variables, the SHG equations can
be reduces to the $C$-integrable Liouville equation, so that the solution
can be expressed explicitly in terms of two arbitrary functions. Some
particular examples of such solutions were studied in \cite{ACD}, and the
Liouville solution was applied to the so-called restricted Cauchy problem
typical for experimental situation in \cite{SFP}.

Here we want to note that the SHG equations (\ref{eq1}) can be approximated
under certain condition by the system which is $C$-integrable and its
solution is given in an explicit form in terms of four arbitrary functions.
This limit is in a sense opposite to the discussed previously pure
modulation case. (Note, however, that a pure modulation case is an exact
reduction of the SHG equations, whereas here we consider some approximation
to these equations.) The proposed approximation is based on the following
observation. Elimination of $q_2$ from Eqs.~(\ref{eq1}) leads to the
equation 
\begin{equation}  \label{eq2}
q_{1,\chi\tau}q_1^*-q_{1,\chi}q_{1,\tau}^*=-2\left(q_1q_1^*\right)^2,
\end{equation}
where the left hand side is quadratic and the right hand side is of fourth
degree in the amplitude $q_1$. Hence, one may suggest that in the limit of
small enough $|q_1|$ we can neglect the right hand side, if derivatives of $%
q_1$ in the left hand side are not too small. To formulate the corresponding
criterion, it is convenient to pass to real variables introduced in \cite{KS}%
.  We represent $q_1$ in the form 
\begin{equation}  \label{eq3}
q_1=\left(\sqrt{Q}/2\right)\exp(i\phi/2),\qquad Q>0,
\end{equation}
so that $q_2$ is given by 
\begin{equation}  \label{eq4}
q_2=-\tfrac14\left[\left(\ln Q\right)_{\chi}+i\phi_{\chi}\right]\exp(i\phi),
\end{equation}
and Eq.~(\ref{eq2}) reduces to the system 
\begin{equation}  \label{eq5}
(\ln Q)_{\chi\tau}-\phi_{\chi}\phi_{\tau}=-Q,\qquad
\left(Q\phi_{\tau}\right)_{\chi}=0.
\end{equation}
Introducing the variables 
\begin{equation}  \label{eq6}
s=1/Q,\quad u=\phi_{\tau},\quad v=(\ln s)_{\chi},\quad w=\phi_{\chi}
\end{equation}
we rewrite the system (\ref{eq5}) in the form \cite{KS} 
\begin{equation}  \label{eq7}
(s)_{\chi}=sv,\quad u_{\chi}=w_{\tau}=vu, \quad v_{\tau}=1/s-uw .
\end{equation}
If we put here $u=0$, we obtain at once the Liouville equation $(\ln
s)_{\chi\tau}=1/s$ which application to SHG was considered in \cite{BS,SFP}.
On the other hand, if $1/s\ll|uw|$, or 
\begin{equation}  \label{eq8}
Q\ll |\phi_{\tau}\phi_{\chi}|,
\end{equation}
we arrive at the system 
\begin{equation}  \label{eq9}
v_{\tau}=-uw, \quad w_{\tau}=uv,\quad u_{\chi}= uv,
\end{equation}
whereas $s$ can be found from the equation $(\ln(s/u))_{\chi}=0$ which
follows from Eqs.~(\ref{eq7}), i.e. 
\begin{equation}  \label{eq10}
s=K(\tau)u,
\end{equation}
where $K(\tau)$ is an arbitrary function.

To find the general solution of the system (\ref{eq9}), we eliminate $v$ and 
$w $, 
\begin{equation}
v=\frac{u_{\chi }}{u},\qquad w=-\frac{1}{u}\left( \frac{u_{\chi }}{u}\right)
_{\tau }=-\frac{u_{\chi \tau }}{u^{2}}+\frac{u_{\chi }u_{\tau }}{u^{3}},
\label{eq11}
\end{equation}
so that $w_{\tau }$ can be presented in the form 
\begin{equation*}
w_{\tau }=-\frac{1}{u}\left( \frac{u_{\tau \tau }-\tfrac{3}{2}({u_{\tau }^{2}%
}/{u})}{u}\right) _{\chi }.
\end{equation*}
According to Eq.~(\ref{eq9}), this must be equal to $u_{\chi }$, hence we
obtain 
\begin{equation*}
\left( \frac{u_{\tau \tau }-\frac{3}{2}({u_{\tau }^{2}}/{u})}{u}\right)
_{\chi }=-\tfrac{1}{2}\left( u^{2}\right) _{\chi },
\end{equation*}
or 
\begin{equation}
u_{\tau \tau }-\tfrac{3}{2}\frac{u_{\tau }^{2}}{u}=-\tfrac{1}{2}u^{3}+f(\tau
)u,  \label{eq12}
\end{equation}
where $f(\tau )$ is an arbitrary function. Generally speaking, this equation
cannot be solved for a given $f(\tau )$ in an explicit form. However, in our
case $f(\tau )$ is an arbitrary function and it can be replaced by another
arbitrary function $F(\tau )$ related with $f(\tau )$ in the following way.
We make a substitution 
\begin{equation}
u=F^{\prime }(\tau )U(\xi ),\qquad \xi =F(\tau ),  \label{eq13}
\end{equation}
so that Eq.~(\ref{eq12}) reduces to 
\begin{equation*}
U_{\xi \xi }-\tfrac{3}{2}({U_{\xi }^{2}}/{U})+\tfrac{1}{2}U^{3}=\frac{1}{%
(F^{\prime }(\tau ))^{2}}\left[ f(\tau )+\frac{3}{2}\left( \frac{F^{\prime
\prime }(\tau )}{f^{\prime }(\tau )}\right) ^{2}-\frac{F^{\prime \prime
\prime }(\tau )}{F^{\prime }(\tau )}\right] U.
\end{equation*}
Thus, if $f(\tau )$ is expressed in terms of $F(\tau )$ by the equation 
\begin{equation}
f(\tau )=\frac{F^{\prime \prime \prime }(\tau )}{F^{\prime }(\tau )}-\frac{3%
}{2}\left( \frac{F^{\prime \prime }(\tau )}{f^{\prime }(\tau )}\right)
^{2}\equiv \left\{ F,\tau \right\} ,  \label{eq14}
\end{equation}
where the curly bracket denotes the Schwarzian derivative \cite{Hille}, then 
$U(\xi )$ is determined by the equation 
\begin{equation}
U_{\xi \xi }-\tfrac{3}{2}({U_{\xi }^{2}}/{U})+\tfrac{1}{2}U^{3}=0.
\label{eq15}
\end{equation}
This equation is solved by elementary methods to give 
\begin{equation*}
U(\xi )=\left[ 1/G+\tfrac{1}{4}(\xi -H)^{2}\right] ^{-1},
\end{equation*}
where $G$ and $H$ are integration constants (arbitrary functions of $\chi $%
). Making use of Eqs.~(\ref{eq13}) and (\ref{eq11}), we arrive at the
general solution of the system (\ref{eq9}): 
\begin{equation}  \label{eq17}
\begin{split}
u(\chi ,\tau )&=\frac{4F^{\prime }G}{4+G^{2}(F-H)^{2}}, \\
v(\chi ,\tau )&=\frac{-GG^{\prime }(F-H)^{2}+2G^{2}H^{\prime
}(F-H)+4G^{\prime }/G}{4+G^{2}(F-H)^{2}}, \\
w(\chi ,\tau )&=\frac{\frac{1}{2}G^{3}H^{\prime }(F-H)^{2}+4G^{\prime
}(F-H)-2GH^{\prime }}{4+G^{2}(F-H)^{2}},
\end{split}%
\end{equation}
where $F(\tau ),$ $G(\chi )$, $H(\chi )$ are arbitrary functions. Together
with Eq.~(\ref{eq10}), these formulas give full description of the SHG
problem in the limit (\ref{eq8}).

>From mathematical point of view, the fact of $C$-integrability of the
system (\ref{eq9}) is not trivial and finds its explanation in possibility
to reduce this system to the so-called modified Liouville equation 
(see Appendix).

One may hope that this solution will permit one to give analytic description
of SHG processes for concrete examples in a way similar to that used in \cite%
{SFP}.

We thank A.V.~Zhiber for indication of Ref.~\cite{Vessiot} and H.~Steudel
for useful remarks on the manuscript. AMK is grateful to DFG (grant 436 RUS
113/89/2 (R,S)) and INTAS (grant 96--0339) for partial support. MVP is
grateful to RFBR (grants 00-01-00210 and 00-01-00366).

\setcounter{equation}{0}

\renewcommand{\theequation}{A.\arabic{equation}}


\subsection*{Appendix}

Let us show that the system (\ref{eq9}) can be reduced to the modified
Liouville equation first introduced by E.~Vessiot \cite{Vessiot} (see also
the paper \cite{SZh}).

We notice that the last two equations (\ref{eq9}) yield a first integral 
\begin{equation}  \label{a1}
v^2+w^2=\left( \psi^{\prime}(\chi)\right)^2,
\end{equation}
where $\psi(\chi)>0$ is an arbitrary function. Then transition to polar
coordinates 
\begin{equation}  \label{a2}
v=\psi^{\prime}(\chi)\cos\theta,\qquad w=\psi^{\prime}(\chi)\sin\theta,
\end{equation}
transforms the last two equations (\ref{eq9}) to 
\begin{equation}  \label{a3}
u=\theta_{\tau},
\end{equation}
and the first equation (\ref{eq9}) takes the form $\theta_{\tau\chi}=\psi^{%
\prime}(\chi)\cos\theta\cdot\theta_{\tau}$ or 
\begin{equation}  \label{a4}
\theta_{\tau\zeta}=\cos\theta\cdot\theta_{\tau},
\end{equation}
where $\zeta=\psi(\chi)$. (This equation can be integrated once and reduced
to a linear second order equation; see Eq.~(\ref{a21}) below.) Introducing a
new variable 
\begin{equation}  \label{a5}
\phi=\ln u=\ln\theta_{\tau},
\end{equation}
we obtain 
\begin{equation*}
\phi_{\zeta}=\theta_{\tau\zeta}/\theta_{\tau}=\cos\theta,\quad
\phi_{\zeta\tau}=-\sin\theta\cdot\theta_{\tau},
\end{equation*}
and finally 
\begin{equation}  \label{a6}
\phi_{\tau\zeta}=-e^{\phi}\sqrt{1-\phi_{\zeta}^2},
\end{equation}
which is just the modified Liouville equation \cite{Vessiot}.
(Amplitude-modulated reduction of degenerate two-photon propagation
equations discussed in \cite{StK} can be also reduced to this equation.)

This observation permits us to write down the solution of Eq.~(\ref{a6}).
Indeed, the general solution (\ref{eq17}) gives 
\begin{equation*}
({\psi ^{\prime }}(\chi ))^{2}=v^{2}+w^{2}=(G^{\prime }/G)^{2}+\tfrac{1}{4}%
(GH^{\prime })^{2}.
\end{equation*}
Hence, if we choose $\psi ^{\prime }(\chi )=1$, so that $\zeta =\chi $ and $%
H(\chi )$ is connected with $G(\chi )$ by the equation 
\begin{equation}
({H^{\prime }})^{2}=\frac{4}{G^{4}}[G^{2}-({G^{\prime }})^{2}],  \label{a7}
\end{equation}
then the solution of Eq.~(\ref{a6}) is given by 
\begin{equation}
\phi =\ln u=\ln \frac{4F^{\prime }G}{4+G^{2}(F-H)^{2}}.  \label{a8}
\end{equation}

Note also that another system (arisen first in a physical problem about
amplification of two counter propagating light beams by transverse flow of
an amplifying medium) 
\begin{equation*}
a_{\tau }=ac,\quad b_{\tau }=-bc,\quad c_{\chi }=-(a+b)c,
\end{equation*}
whose general solution was found long ago \cite{KCh}, can be reduced by
substitutions 
\begin{equation*}
u=c,\quad w=b-a,\quad v=-(a+b),
\end{equation*}
to the system 
\begin{equation}  \label{a9}
v_{\tau }=uw,\quad w_{\tau }=uv,\quad u_{\chi }=uv,
\end{equation}
which differs from Eqs.~(\ref{eq9}) only by the sign in the right hand side
of the first equation, and is connected with another forms of the modified
Liouville equation 
\begin{equation}  \label{a10}
\phi _{\tau \zeta }=e^{\phi }\sqrt{\phi _{\zeta }^{2}\pm 1},
\end{equation}
depending on the choice of the sign in the first integral 
\begin{equation}  \label{a11}
v^{2}-w^{2}=\mp \left( \psi ^{\prime }(\chi )\right) ^{2}.
\end{equation}

General solution of Eqs.~(\ref{a9}) is given by the formulas 
\begin{equation}  \label{a12}
\begin{split}
u(\chi ,\tau )& =\frac{4F^{\prime }G}{4-G^{2}(F-H)^{2}}, \\
v(\chi ,\tau )& =\frac{GG^{\prime }(F-H)^{2}-2G^{2}H^{\prime
}(F-H)+4G^{\prime }/G}{4-G^{2}(F-H)^{2}}, \\
w(\chi ,\tau )& =\frac{-\frac{1}{2}G^{3}H^{\prime }(F-H)^{2}+4G^{\prime
}(F-H)-2GH^{\prime }}{4-G^{2}(F-H)^{2}}.
\end{split}%
\end{equation}
Then we have 
\begin{equation*}
v^{2}-w^{2}={G^{\prime ^{2}}}/{G^{2}}-\tfrac14{G^{2}H^{\prime ^{2}}},
\end{equation*}
and 
\begin{equation}  \label{a13}
\phi=\ln\frac{4F^{\prime}G}{4-G^2(F-H)^2}
\end{equation}
is the solution Eqs.~(\ref{a10}) provided $H$ satisfies the equation 
\begin{equation}  \label{a14}
H^{\prime ^{2}}={4}[G^{\prime ^{2}}\pm G^{2}]/{G^{4}}
\end{equation}
with a proper choice of the sign.

As was noted in \cite{Vessiot}, solutions of the modified Liouville
equations (\ref{a6}) are connected with solutions of the Liouville equation 
\begin{equation}  \label{a15}
z_{\tau \zeta }=e^{z}
\end{equation}
by differential substitutions 
\begin{equation}  \label{a16}
z=\phi +\ln(\phi_{\zeta }+\sqrt{\phi_{\zeta}^2\pm 1}), 
\end{equation}
which can be written in terms of hyperbolic functions as follows, 
\begin{equation}  \label{a17}
z=\phi +\mathrm{Arsinh}\phi _{\zeta } \quad \text{or}\quad z=\phi +\mathrm{%
Arcosh}\phi _{\zeta }.
\end{equation}
These substitutions map the general solutions (\ref{a13},\ref{a14}) of the
modified Liouville equations (\ref{a10}) into the general solution 
\begin{equation}  \label{a18}
z=\ln \frac{2F^{\prime }(\tau )Q^{\prime }(\zeta )}{(F(\tau)+Q(\zeta))^{2}},
\end{equation}
of the Liouville equation (\ref{a15}), where 
\begin{equation}  \label{a19}
Q(\zeta)=-2/G(\zeta)-H(\zeta).
\end{equation}

Substitutions (\ref{a17}) can be reduced to a second order linear equation
for the variable $y$ defined by $y_{\zeta}/y=\tfrac12\exp(z-\phi)$ which is
considered as a function only of $\zeta$: 
\begin{equation}  \label{a21}
y_{\zeta\zeta}-\left(\frac{Q^{\prime\prime}}{Q^{\prime}}-\frac{2Q^{\prime}}{%
F+Q}\right)y_{\zeta}\mp\tfrac14 y=0.
\end{equation}
The above consideration gives one solution of this second order equation in
the form 
\begin{equation}  \label{a22}
y=\exp\left(\tfrac12\int\exp(z-\phi)d\zeta\right).
\end{equation}

At last we notice that Eq.~(\ref{a6}) does not admit real substitution (\ref%
{a17}) (but only complex one) which transforms this equation into Liouville
equation. Nevertheless, if we write the solution of the modified Liouville
equation 
\begin{equation}  \label{a23}
\phi _{\tau \zeta }=-e^{\phi }\sqrt{1-\varepsilon ^{2}\phi _{\zeta }^{2}}
\end{equation}
in the form 
\begin{equation}  \label{a24}
\phi =\ln \frac{2F^{\prime }(\tau )G(\zeta )}{(F(\tau)-H(\zeta))^{2}+
\varepsilon ^{2}G^{2}(\zeta)},
\end{equation}
where 
\begin{equation*}
H^{\prime ^{2}}=G^{2}-\varepsilon ^{2}G^{\prime ^{2}},
\end{equation*}
then in the limit $\varepsilon\to 0$ it goes directly to the solution of the
Liouville equation $\phi_{\tau\zeta}=-\exp\phi$.

\end{document}